\documentclass[12]{article}
\usepackage{epsfig,amsmath,amssymb,enumitem}
\usepackage[papersize={8.5in,11in}]{geometry}
\usepackage{multirow}
\begin{document}
\title{Quarkonium Resonance Model of the 125 GeV Boson}
\author{J. W. Moffat\\~\\
Perimeter Institute for Theoretical Physics, Waterloo, Ontario N2L 2Y5, Canada\\
and\\
Department of Physics and Astronomy, University of Waterloo, Waterloo,\\
Ontario N2L 3G1, Canada}
\maketitle
\begin{abstract}
A quarkonium model of the recently discovered X boson at 125 GeV is investigated.  Two isosinglet $J^{PC}=0^{-+}$ resonances $\zeta$ and $\zeta'$ mix and their masses are determined by the masses of the bottomonium and toponium eigenstates. This mixing is enhanced by the non-perturbative gluon interaction produced by an axial $U_A(1)$ anomaly. The resonance $\zeta$ is identified with the 125 GeV boson X, while the heavier $\zeta'$ resonance has a mass 230 GeV. We assess the most recent data for the decay mode $X\rightarrow ZZ^*\rightarrow$ 4 leptons with a critical discussion of the determination of the parity of the $X$ boson.
\end{abstract}

\section{Introduction}

The discovery of the new boson at an energy $125$ GeV at the Large Hadron Collider (LHC) has been heralded as the discovery of the long-sought standard model Higgs boson. The latest CMS~\cite{CMS} and ATLAS~\cite{ATLAS} experiments based on $5\,{\rm fb}^{-1}$ data at 7 TeV and $12-20 \,{\rm fb}^{-1}$ data at 8 TeV for the decay of the new boson $X\rightarrow\gamma\gamma$ and $X\rightarrow ZZ^*\rightarrow 4$ leptons strengthen the case for the identification of the new boson X with the standard model Higgs boson. However, there are certain discrepancies in the data collected so far with the predictions of the Higgs boson, such as weak or zero signals for $X\rightarrow WW^*\rightarrow \nu\bar\nu\ell^+\ell^-$, $X\rightarrow \tau^+\tau^-$ and $b\bar b$. Moreover, there are insufficient data in the decay $X\rightarrow ZZ^*\rightarrow 4\ell$ to discriminate between a scalar X(125) and a pseudoscalar X(125) boson~\cite{Eichten,Moffat}. The multivariate analyses discriminating a scalar and pseudoscalar X(125) perfomed by CMS and ATLAS are model dependent~\cite{Choi,Gao,Rujula,Logan,Ellis,Stolarski,Gao2,Drozdetskiy,Djouadi}. Only effective higher-dimensional pseudoscalar Higgs boson models have been compared to the standard model scalar Higgs boson model~\cite{ATLAS1,CMS1}.

Without the discovery of new physics beyond the standard model, the standard model with an elementary Higgs boson suffers from serious theoretical problems, such as the Higgs mass hierarchy problem, the gauge hierarchy problem associated with the electroweak and Planck scales,  and the number, mixing and flavor problems in the fermion sector.

Because the new boson X decays into two photons, an experimental signature which is considered a decisive confirmation of its existence, then according to the Landau-Yang theorem~\cite{Landau,Yang} the X boson must have either spin 0 or 2. Because a "graviton-like" spin 2 boson seems an unlikely candidate for the X boson, then spin 0 is considered as the most plausible choice. However, all the observed particles of the standard model have either spin 1/2 or 1. The only {\it elementary} particle that can fill the role of a spin 0 boson is the standard model Higgs boson with $J^{PC}=0^{++}$. But the new boson X could be a composite particle. Quark-antiquark composite states and their spectroscopy abound in experimental particle physics, including excited states with $J^{PC}=0^{++}$~\cite{pdg}. An alternative explanation for the X boson is that it is a composite particle. A recent paper argues that it is a techni-pion with spin 0 and $PC=-1$~\cite{Eichten}. Another possibility is that it is a mixed quark-antiquark resonance called $\zeta^0$ with $J^{PC}=0^{-+}$ and a ground state dominant $^1S_0$ decay into two photons ~\cite{Moffat2}. Such a bound state system is analogous to positronium with a dominant decay into two photons.

This identification of the $X$ boson implies that it is not an elementary Higgs boson. If this is true, what can we say about electroweak gauge theory? Possible Higgless models with dynamical gauge symmetry breaking can correctly predict the W and Z boson masses and account for the quark and lepton mass spectrum~\cite{Moffat3,Moffat4,Moffat5}. Other possibilities exist in the literature.

The effective constituent quark mass of the $\zeta$ boson is $m_{q{\rm eff}}\sim m_\zeta/2\sim 62-63$ GeV. This quark mass is used to calculate the decay rates of the $\zeta(125)$ boson and the associated branching ratios. A detailed derivation of the partial decay widths and branching ratios of the $\zeta$ boson has been compared to the predicted decay widths of the standard model Higgs boson~\cite{Moffat2}. The branching ratios for the $\zeta$ decays into bosons are comparable in size to the Higgs boson decays, but the fermion-anti-fermion decay rates are suppressed compared to the Higgs boson. An excess of the signal strength of the two photon decay of the $\zeta$ meson can be obtained from calculated production cross sections and branching ratios, which is consistent with the current CMS and ATLAS data~\cite{CMS,ATLAS}. An analysis of the angular distributions in the decay $X\rightarrow ZZ^*\rightarrow 4\ell$ that can discriminate the elementary scalar Higgs boson from the pseudoscalar quarkonium $\zeta$ boson remains to be performed to test the true quantum numbers of the X boson.

\section{The $\zeta$ and $\zeta'$ resonances as mixtures of bottomonium and toponium}

With respect to an effective interaction Hamiltonian, heavy quarkonium resonances appear in two different isoscalar states $\vert\zeta^0\rangle$ and $\vert\zeta^{0'}\rangle$. The effective Hamiltonian is given by~\cite{Moffat2,Moffat}:
\begin{equation}
{\cal H}_{\rm eff}={\cal H}_0+{\cal H}_{\rm mass},
\end{equation}
where
\begin{equation}
{\cal H}_{\rm mass}=K^T{\cal M}K.
\end{equation}
Here, ${\cal M}$ is the mass matrix:
\begin{equation}
\label{Massmatrix}
{\cal M}=\biggl(\begin{array}{cc}m_{\zeta'}&m_{\zeta\zeta'}\\
m_{\zeta\zeta'}&m_{\zeta}\end{array}\biggr),
\end{equation}
where $\vert\zeta\rangle$ and $\vert\zeta'\rangle$ are states of quarkonium that interact through the mixing contributions $m_{\zeta\zeta'}$ and $K=\biggl(\begin{array}{cc}\zeta'\\\zeta\\\end{array}\biggr)$.
The mass matrix can be diagonalized:
\begin{equation}
\label{Diagonalmatrix}
D=R^T{\cal M}R,
\end{equation}
where $R$ is the rotation matrix:
\begin{equation}
\label{rotationmatrix}
R=\biggl(\begin{array}{cc}\cos\phi&\sin\phi\\-\sin\phi&\cos\phi\end{array}\biggr),
\end{equation}
and $D$ is the diagonal matrix:
\begin{equation}
D=\biggl(\begin{array}{cc}m_T&0\\0&m_B\end{array}\biggr),
\end{equation}
where the observed bottomonium and unobserved toponium are isoscalar states $\vert B\rangle=\vert b{\bar b}\rangle$ and $\vert T\rangle=\vert t{\bar t}\rangle$ of heavy quarkonium. Here, $m_T\sim 2m_t\sim 346$ GeV and $m_B\sim 2m_b\sim 9$ GeV where we have used the measured quark masses: $m_t\sim 173$ GeV and $m_b\sim 4.5$ GeV~\cite{pdg}.

In terms of the states $\vert T\rangle$ and $\vert B\rangle$, we have
\begin{equation}
{\cal H}'_{\rm mass}=m_TTT^\dagger+m_BBB^\dagger,
\end{equation}
where
\begin{equation}
\biggl(\begin{array}{cc}T\\B\\\end{array}\biggr)
=R\biggl(\begin{array}{cc}\zeta'\\\zeta\\\end{array}\biggr)
=\biggl(\begin{array}{cc}\cos\phi \zeta'+\sin\phi \zeta\\\cos\phi\zeta-\sin\phi\zeta'\end{array}\biggr).
\end{equation}
From (\ref{Diagonalmatrix}), we obtain the masses:
\begin{equation}
m_T=\cos^2\phi m_{\zeta'}-2\sin\phi\cos\phi m_{\zeta\zeta'}+\sin^2\phi m_\zeta,
\end{equation}
and
\begin{equation}
m_B=\cos^2\phi m_\zeta+2\sin\phi\cos\phi m_{\zeta\zeta'}+\sin^2\phi m_{\zeta'}.
\end{equation}

By inverting (\ref{Diagonalmatrix}), we get
\begin{equation}
{\cal M}=RDR^T
\end{equation}
which leads to the result
\begin{equation}
\label{zetasolution}
{\cal M}\equiv\biggl(\begin{array}{cc}m_{\zeta'}&m_{\zeta\zeta'}\\
m_{\zeta\zeta'}&m_\zeta\end{array}\biggr)=\biggl(\begin{array}{cc}\cos^2\phi m_T+\sin^2\phi m_B&\cos\phi\sin\phi(m_B-m_T)\\
\cos\phi\sin\phi(m_B-m_T)&\cos^2\phi m_B+\sin^2\phi m_T\end{array}\biggr).
\end{equation}
We derive from (\ref{zetasolution}) the results
\begin{equation}
\label{zetamass}
m_\zeta=\cos^2\phi m_B+\sin^2\phi m_T,
\end{equation}
and
\begin{equation}
\label{zetaprimemass}
m_{\zeta'}=\cos^2\phi m_T+\sin^2\phi m_B.
\end{equation}
The off-diagonal term is given by
\begin{equation}
m_{\zeta\zeta'}=\cos\phi\sin\phi(m_B-m_T).
\end{equation}

The mixing angle $\phi$ is obtained from the equation:
\begin{equation}
\label{mixingangle}
\phi=\arccos[(m_T-m_{\zeta})/(m_T-m_B)]^{1/2}.
\end{equation}
The mixing angle $\phi=36\,^{\circ}$ is obtained from (\ref{mixingangle}) and from (\ref{zetamass}) and (\ref{zetaprimemass}), we get the masses of the quarkonium states $\vert\zeta\rangle$ and $\vert\zeta'\rangle$: $m_{\zeta^0}\sim 125$ GeV and $m_{\zeta^{0'}}\sim 230$ GeV. We identify the new boson $X(125)$ discovered at the LHC with the $\zeta^0$ bound state quarkonium resonance.  The strength of the mixing between $\zeta$ and $\zeta'$ is given by
\begin{equation}
m_{\zeta\zeta'}=\cos\phi\sin\phi(m_B-m_T)=-160\,{\rm GeV}.
\end{equation}
In the next section, we will justify the strength of the $\zeta^0$ and $\zeta^{0'}$ mixing and the mixing angle $\phi$ by the axial $U_A(1)$ anomalous non-perturbative gluon interaction between the isosinglet $\zeta^0$ and $\zeta^{0'}$ mesons.

We have used the zero width approximation to determine the diagonalization of the mass matrix and the masses of the $\zeta$ and $\zeta'$ resonances. We could have determined the mixing of the mass matrix using the finite resonance width or overlapping resonance formalism~\cite{Dothan,Dothan2,Gilman}. The $2\times 2$ mass matrix then takes the form:
\begin{equation}
\label{BRmassmatrix}
{\cal M}_0=\biggl(\begin{array}{cc}m_{0\zeta'}-i\Gamma_{0\zeta'}&\delta m\\
\delta m&m_{0\zeta}-i\Gamma_{0\zeta}\end{array}\biggr).
\end{equation}
Here, ${\cal M}_0$ is the undiagonalized mass matrix with elements expressed in terms of bare masses and widths $m_0$ and $\Gamma_0$ and the off-diagonal term $\delta m$, which induces the mixing, originates in the interaction of the $\zeta$ and $\zeta'$ mesons.

\section{Mixing strength of the $\zeta$ and $\zeta'$ bosons}

Quantum chromodynamics (QCD) has, in addition to $SU(N_f)\times SU(N_f)$ chiral symmetry, an approximate $U_A(1)$ symmetry. In the latter symmetry, all left-handed quark fields are rotated by a common phase while the right-handed quark fields are rotated by the opposite phase. The $U_A(1)$ symmetry is violated by the quantum field theory axial anomaly and cannot produce a Goldstone boson when $U(N_f)\times U(N_f)$ chiral symmetry is spontaneously broken. The $\eta'(958)$ particle mass for the $N_f=3$ case, acquires a large mass through the instanton quantum tunneling effect, which breaks the mass degeneracy with pions, kaons, and the pseudoscalar $\eta'$ meson in the chiral limit when all quarks are massless. The vacuum breaking of the conserved axial currents should imply the breaking of symmetries which these currents generate, and in turn this should imply the existence of nine Goldstone bosons. However, this conclusion is inconsistent with the large masses of the isosingle $\eta$ and $\eta'$ mesons fomed from the very light $u, d$ and $s$ quarks, and the large splitting between the 8 octet mesons, $\pi^{+-}, \pi^0, K^{+-}, K^0,\bar {K}^0,\eta^0$ and the singlet $\eta^{0'}$. The $U_A(1)$ axial current corresponding to the singlet $\eta'$ meson is unique in possessing an anomalous divergence at the quantum level. The instanton configuration with non-trivial gluon $Q_\eta$ charge topology was discovered as a mechanism to explain the large $\eta'$ mass~\cite{Polyakov,tHooft,Witten,Veneziano,Gross,Wilczek,Huang,Kapusta,Christ,Shore,Shore2}.

The mass scale ratio between the $\eta(958)$ and the $\eta'(548)$ is 1.75, while for the $\zeta(125)$ and the $\zeta'(230)$ it is 1.84, which strongly suggests that the strength of interaction and mixing of the $\zeta$ and $\zeta'$ will also be dominated by an axial current $U_A(1)$ topological gluon charge
$Q_\zeta$. Indeed, the perurbative QCD gluon interaction is too weak to produce a mixing angle as large as $\phi=36\,^{\circ}$.  We postulate a mass degeneracy between the heavy $\zeta$ and $\zeta'$ mesons with the chiral symmetry $SU(2)\times SU(2)$. The quark part of the QCD action is invariant under the global flavor symmetry $G_f=U_A(1)\times SU(2)\times SU(2)$, and the $U_A(1)$ symmetry valid classically can be violated by quantum effects. The symmetry is violated by the non-conservation of the quark axial current $J_\mu^5=\bar q\gamma_\mu\gamma^5q$~\cite{Adler,Bell}:
\begin{equation}
\partial_\mu J^{5\mu}=\frac{N_f\alpha_s}{4\pi}Tr\biggl(G_{\mu\nu}\tilde{G}^{\mu\nu}\biggr),
\end{equation}
where $\alpha_s$ is the strong interaction coupling constant and $G^{\mu\nu}$ is the gluon field strength. The gluon topology can be formulated in terms of the gluon topological susceptability
\begin{equation}
\chi(0)=i\int d^4x\langle 0\vert TQ(x)Q(0)\vert 0\rangle,
\end{equation}
where
\begin{equation}
Q=\frac{\alpha_s}{4\pi}Tr(G_{\mu\nu}\tilde{G}^{\mu\nu})
\end{equation}
is the gluon topological charge density. The topological susceptability is a correlation function in QCD and is a basic tool for understanding the dynamics of the $U_A(1)$ channel. Non-perturbative studies have been carried out using lattice gauge theory, sum rules and models of the instaton vacuum. The divergences of the flavor singlet axial vector currents associated with the $\zeta$ and $\zeta'$ states acquire anomalous parts, due to the non-perturbative interaction with gluon fields, which do not vanish in the chiral $\zeta$ and $\zeta'$ mass degenerate limit $m_\zeta=m_{\zeta'}$.

We can define the $\zeta$ and $\zeta'$ decays into vector bosons by the amplitudes
\begin{equation}
\langle VV\vert\zeta\rangle=-ig_{\zeta VV}\epsilon_{\lambda\rho\alpha\beta}p_1^\alpha p_2^\beta \epsilon^\lambda(p_1)\epsilon^\rho(p_2)
\end{equation}
and
\begin{equation}
\langle VV\vert\zeta'\rangle=-ig_{\zeta' VV}\epsilon_{\lambda\rho\alpha\beta}p_1^\alpha p_2^\beta \epsilon^\lambda(p_1)\epsilon^\rho(p_2).
\end{equation}
We then obtain for the axial vector currents
\begin{equation}
\langle 0\vert J^3_{5\mu}\vert\zeta^0\rangle=ik_\mu f_\zeta,
\end{equation}
and
\begin{equation}
\langle 0\vert J^3_{5\mu}\vert\zeta^{0'}\rangle=ik_\mu f_{\zeta'}.
\end{equation}
We now get the axial current non-conservation equation
\begin{equation}
\label{nonconservation}
\partial^\mu J^a_{5\mu}=M_{ab}\phi^b_5+\sqrt{2n_f}Q\delta_{a0},
\end{equation}
where $J^a_{5\mu}=\bar q\gamma_\mu\gamma_5 T^aq$ and $\phi^a_5=\bar q\gamma_5T^a q$. The second contribution in (\ref{nonconservation}) is the axial anomaly non-perturbative gluon contribution.

The $\zeta$ and $\zeta'$ decay constants $f^{a\alpha}$ are defined by
\begin{equation}
f^{a\alpha}=\biggl(\begin{array}{cc}f^{0\zeta'}&f^{0\zeta\zeta'}\\
f^{0\zeta\zeta'}&f^{0\zeta}\end{array}\biggr),
\end{equation}
where $a$ is the $SU(2)\times U(1)$ flavor index, and $\alpha$ denotes the physical particle state.

\section{Effective Lagrangians for $\zeta$ boson decays}

An effective Lagrangian for the coupling of the $\zeta$ boson to two photons is given by~\cite{Lansberg}:
\begin{equation}
{\cal L}_{\gamma\gamma{\rm eff}}
=-ig_{\zeta\gamma\gamma}J^5_{q\sigma}\epsilon^{\mu\nu\rho\sigma}F_{\mu\nu}A_\rho=-ig_{\zeta\gamma\gamma}(\bar{q}\gamma_\sigma\gamma^5q)\epsilon^{\mu\nu\rho\sigma}F_{\mu\nu}A_\rho.
\end{equation}
Here, the coupling constant $g_{\zeta\gamma\gamma}$ is
\begin{equation}
g_{\zeta\gamma\gamma}\sim \frac{e_q^2(4\pi\alpha)}{m_Q^2+b_Qm_Q},
\end{equation}
where $e_q$ is the quark charge per unit proton charge and $b_Qm_Q$ is the binding energy of the quarkonium state. The factor $1/(m_Q^2+b_Qm_Q)$ arises from the quark propagator:
\begin{equation}
\Delta_q=\frac{1}{(p_1-p_2)^2/4-m_q^2},
\end{equation}
where $p_1,p_2$ are the outgoing photon momenta.

The effective Lagrangian for the $\zeta$ boson decay into two leptons is~\cite{Lansberg}:
\begin{equation}
{\cal L}_{\ell\ell{\rm eff}}=- g_{\zeta\ell\ell}J_q^{\mu}(\bar\ell\gamma_\mu\ell)=- g_{\zeta\ell\ell}(\bar q\gamma^\mu q)(\bar\ell\gamma_\mu\ell),
\end{equation}
where
\begin{equation}
g_{\zeta\ell\ell}=\frac{e_q(4\pi\alpha)}{m_\zeta^2}.
\end{equation}
The amplitude for the two-lepton decay is given by
\begin{equation}
{\cal M}_{\ell\ell}=e_q(4\pi\alpha)\frac{f_\zeta}{m_\zeta}\epsilon^{\mu}(\bar\ell\gamma_\mu\ell).
\end{equation}
The two-lepton decay width for the $\zeta$ boson is (neglectic lepton masses):
\begin{equation}
\Gamma_{\ell\ell}(\zeta)=\frac{1}{64\pi^2m_\zeta}\int d\Omega\vert{\cal M}_{\ell\ell}\vert^2=\frac{4\pi e^2_q\alpha^2f_\zeta^2}{3m_\zeta}.
\end{equation}

We define
\begin{equation}
\langle 0\vert J^{5\mu}_q\vert Q\rangle=\langle 0\vert\bar q\gamma^\mu\gamma^5\vert Q\rangle=if_Qk^\mu,
\end{equation}
where $f_Q$ is the quarkonium decay constant. We obtain the following expression for the amplitude for $Q\rightarrow \gamma\gamma$:
\begin{equation}
{\cal M}_{\gamma\gamma}=-4ie_q^2(4\pi\alpha)\frac{f_Q}{m_Q^2+b_Qm_Q}\epsilon^{\mu\nu\rho\sigma}
\epsilon_{1\mu}\epsilon_{2\nu}p_{1\rho}p_{2\sigma}.
\end{equation}
This yields the partial $\zeta(^1S_0)$ two-photon decay rate with $b_\zeta\sim 0$:
\begin{equation}
\Gamma(\zeta\rightarrow\gamma\gamma)=\frac{1}{2}\frac{1}{64\pi^2m_\zeta}\int d\Omega\vert{\cal M_{\gamma\gamma}}\vert^2=\frac{4\pi e_q^4\alpha^2f_\zeta^2}{m_\zeta},
\end{equation}
where the factor $1/2$ is the Bose symmetry factor. If we use $f_\zeta^2=12\vert R_S(0)\vert^2/m_Q$, we recover the non-relativistic result of Novikov et al., Barger et al., and Kwong et al.~\cite{Novikov,Barger,Kwong} for the two-photon decay rate.

The effective Lagrangian for the decay of $\zeta$ into $ZZ$ is given by
\begin{equation}
\label{ZZdecay}
{\cal L}_{ZZ{\rm eff}}
=-ig_{\zeta ZZ}J^5_{q\sigma}\epsilon^{\mu\nu\rho\sigma}Z_{\mu\nu}Z_\rho=-ig_{\zeta ZZ}(\bar{q}\gamma_\sigma\gamma^5q)\epsilon^{\mu\nu\rho\sigma}Z_{\mu\nu}Z_\rho,
\end{equation}
where $Z_{\mu\nu}=\partial_\mu Z_\nu-\partial_\nu Z_\mu$. Similarly, the effective Lagrangian for the decay into $WW$ is
\begin{equation}
\label{WWdecay}
{\cal L}_{WW{\rm eff}}
=-ig_{\zeta WW}J^5_{q\sigma}\epsilon^{\mu\nu\rho\sigma}W_{\mu\nu}W_\rho=-ig_{\zeta WW}(\bar{q}\gamma_\sigma\gamma^5q)\epsilon^{\mu\nu\rho\sigma}W_{\mu\nu}W_\rho,
\end{equation}
where $W_{\mu\nu}=\partial_\mu W_\nu-\partial_\nu W_\mu$. For the $\zeta$ decay into $ZZ,WW$ one of the $Zs$ and one of the $Ws$ is off the mass shell, because the $\zeta(125)$ mass is below the $ZZ,WW$ threshold.

In the current literature, the multivariate analyses of spin and parity of the new $X$ boson are model dependent. The Higgs boson contribution to the standard model Lagrangian is determined by the gauge invariant expression:
\begin{equation}
\label{gaugederivation}
(D^\mu\phi)(D_\mu\phi)^\dagger,
\end{equation}
where $D_\mu$ is the covariant derivative involving the $B^\mu$ and $W^{i\mu}$ boson fields. The effective Lagrangian for the pseudoscalar Higgs boson cannot be derived from the standard model gauge invariant Lagrangian formalism using (\ref{gaugederivation}). The model dependent effective Lagrangian for the pseudoscalar Higgs boson takes the form:
\begin{equation}
\label{pseudoscalarHiggs}
{\cal L}_{PS{\rm eff}}=g_{PS1}\frac{\alpha_s}{4\pi v}\phi_{PS}G^a_{\mu\nu}\tilde{G}^{a\mu\nu}
+g_{PS2}\frac{\alpha}{4\pi v}\phi_{PS}(B_{\mu\nu}\tilde{B}^{\mu\nu}+g_{PS3}W^i_{\mu\nu}\tilde{W}^{i\mu\nu}),
\end{equation}
where $v=\langle 0\vert\phi\vert 0 \rangle=$ 246 GeV is the vacuum expectation value of the scalar Higgs field, and $\tilde{G}^{a\mu\nu}=
\epsilon^{\mu\nu\rho\sigma}G^a_{\rho\sigma}$.
These effective pseudoscalar Lagrangians describe dimension-5 operator Lagrangians which are not renormalizable.

In momentum space, the most general $HVV$ vertex can be written: $V_{\mu\nu}(p,q)
=V_{\mu\nu}^{SM}(p,q)+V_{\mu\nu}^{BSM}(p,q)$ where we parameterize the coupling as $iV^{\mu\nu}(p,q)\epsilon_\mu(p)\epsilon_{\nu}^*(q)$. We have
\begin{equation}
V^{SM}_{\mu\nu}=-gm_V\eta_{\mu\nu},
\end{equation}
\begin{equation}
\label{couplingmomentum}
V^{BSM}_{\mu\nu}(p,q)=\frac{g}{m_V}(h(p\cdot q\eta_{\mu\nu}-p_\mu p_\nu)+h'\epsilon_{\mu\nu\rho\sigma} p^\rho q^\sigma).
\end{equation}
Here, $\eta_{\mu\nu}$ is the Minkowski metric and $h$ and $h'$ are effective coupling strengths for $PC=+1$ and $PC=-1$ couplings, respectively.

The calculation of the decay rates of the pseudoscalar Higgs boson are not direct tree graph calculations.  The calculated decay rates are significantly weakened perturbatively by occurring through loops. Therefore, the phenomenological coupling constants in  (\ref{pseudoscalarHiggs}) and ({\ref{couplingmomentum}) must be boosted in an {\it ad hoc} manner to give reasonable decay rates comparable to the standard model Higgs boson decay rates and branching ratios. This will introduce a significant bias into the angular distribution spin and parity analyses of the decay channels $H\rightarrow ZZ^*\rightarrow 4\ell$, $H\rightarrow WW^*\rightarrow \nu\nu\ell\ell$.  Apart from this issue, the present data as of December 2012 for the $H\rightarrow ZZ^*\rightarrow 4\ell$ consisting of 16 events for the CMS result and 17-18 events for the ATLAS result are insufficient to discriminate between the positive and negative parity of the $X$ boson.\begin{footnote}{ Eichten et al.,~\cite{Eichten} give a critical analysis of the latest CMS and ATLAS data for the decay channels $H\rightarrow ZZ^*\rightarrow 4\ell$ and $H\rightarrow WW^*\rightarrow \nu\bar\nu\ell^+\ell^-$.  They conclude that the CMS and ATLAS results suffer from a deficiency of real $Z$ and $W$ events to be able to make a fully convincing claim for the $X$ boson being the scalar Higgs boson. Further data analysis based on the 2012 LHC run will hopefully clarify this situation.}\end{footnote}

To discriminate between the scalar Higgs boson parity $PC=+1$ and the quarkonium resonance model pseudoscalar parity $PC=-1$, we must compare the standard model Higgs Lagrangian with the the quarkonium resonance effective Lagrangians (\ref{ZZdecay}) and (\ref{WWdecay}). The latter Lagrangians are equivalent to tree graph Feynman diagrams as in the case of the Higgs boson, so comparable decay rates to $ZZ^*$ and $WW^*$ are to be expected to be predicted by the quarkonium resonance model~\cite{Moffat2}. This is in contrast to the pseudoscalar Higgs boson decays into $ZZ^*$ and $WW^*$. Only a comparison of the scalar Higgs model with a non-Higgs model, such as the quarkonium resonance model, can be expected to produce a physically significant way to discriminate between the positive and negative parities of the $X$ boson. Both models predict non-vanishing and potentially statistically significant $X\rightarrow ZZ^*\rightarrow 4 \ell$ and $X\rightarrow WW^*\rightarrow \nu\nu\ell\ell$ decay rates providing reliable spin-parity analyses.

An {\it effective} constituent quark mass, $m_{q{\rm eff}}\sim m_\zeta/2\sim 62-63$ GeV, and an effective quark charge for an up-type quark, $e_q=(2/3)e$, was used to calculate the decay rates of the $\zeta(125)$ boson and the associated branching ratios~\cite{Moffat2}. The decay rates and branching ratios of the $\zeta$ boson were calculated and compared to the predicted Higgs boson decay rates and branching ratios.

The decay rate and branching ratio of the $\zeta$ boson to $\gamma\gamma$ can be fitted to the ATLAS and CMS data, while the decay rates and branching ratios to $ZZ^*, WW^*$ and $Z\gamma$ are comparable to the decay rates predicted by the standard model Higgs boson. The decay rates of $\zeta$ to the $\tau^+ \tau^-$ leptons and the $b\bar b$ and $c\bar c$ quarks are suppressed.  This constitutes a crucial test of the quarkonium resonance model.

Due to the large background from the decay $Z\rightarrow\tau^+\tau^-$, the decay $X\rightarrow\tau^+\tau^-$ can be detected in $WX\rightarrow\ell\nu\tau\tau$ and $ZX\rightarrow\ell^+\ell^-\tau\tau$. In July, the CMS reported the signal strength $\mu(\tau^+\tau^-)=0.0\pm 0.8$ and in November $\mu(\tau^+\tau^-)=0.9\pm 0.5$. The results for $W/ZX$ production are consistent with a Higgs boson decay, but with large errors. In November, ATLAS reported a signal strength $\mu(\tau^+\tau^-)=0.8\pm 0.8$. The reported signal strengths by both the CMS and ATLAS collaborations are weak, which is not surprising considering the large background. The shift in the signal strength $\mu(\tau^+\tau^-)$ could be due to statistical fluctuations or a problem with the analysis of the background. In November, CMS reported $\mu(b\bar b)=1.1\pm 0.6$ from the $WX\rightarrow \ell\nu b\bar b$ channel, while ATLAS has reported no signal, $\mu(b\bar b)=-0.4\pm 1.0$. The combined Tevatron CDF and D0 experiments claimed a $3.1\sigma$ statistical significance for $H\rightarrow\tau^+\tau^-$ even though they had a weak $S/B$ signal~\cite{Aaltonen}. The broad excess of events was not a convincing fit to the $X(125)$ boson with a more significant excess at 135 GeV. As with the $\tau^+\tau^-$ channel the excessive background for the $b\bar b$ decay channel causes the difficulty in detecting a positive signal.  Using the full Tevatron CDF Run II data set, a search for the standard model Higgs boson based on Higgs boson production from gluon-gluon fusion, vector-boson fusion, and associated production with either a $W$ or $Z$ boson or a $t\bar t$ pair, claimed a 2$\sigma$ excess of Higgs boson events~\cite{Aaltonen2}. They reported a signal strength cross section for $H\rightarrow\tau^+\tau^-$ for $X(125)$ of magnitude $0.0^{+8.44}_{-0.0}$ which is compatible with zero. We must await further data analyses to decide whether the decay of the $X$ boson to fermi-antifermi pairs is in fact supressed.

\section{Conclusions}

We have developed a model in which the newly discovered boson at the LHC with a mass $125-126$ GeV can be identified with a heavy quarkonium, spin 0 pseudoscalar resonance $\zeta^0$. By mixing the two states $\vert\zeta\rangle$ and $\vert\zeta'\rangle$ through a rotation angle $\phi\sim 36\,^{\circ}$, we obtain two heavy quarkonium states $\vert\zeta\rangle$ and $\vert\zeta'\rangle$ and the mass of the $\zeta'$ is $m_{\zeta'}\sim 230$ GeV. The $\zeta^0$ boson is a bound state with QCD gluon interactions. The strength of the $\zeta$ and $\zeta'$ meson mixing and the size of the mixing angle are significantly enhanced by the non-perturbative $U_A(1)$ anomaly of the quark-antiquark axial currents.

The $\zeta$ boson decay rates into bosons and fermion-antifermion states and their branching ratios are calculated using non-relativistic, heavy quarkonium decay formulas with an effective constituent quark mass, $m_{q{\rm eff}}\sim m_\zeta/2\sim 62-63$ GeV~\cite{Moffat2}. With sufficient precise data for the decay rates and the branching ratios of the X boson, it will eventually be possible to discriminate between the quarkonium spin zero resonance $\zeta$ and the standard model Higgs boson. In particular, with enough data it will be possible to determine whether the X boson is a scalar or pseudoscalar boson. Effective Lagrangians for the quarkonium resonance $\zeta$ coupled to $\ell\ell$, $\gamma\gamma$, $ZZ^*$ and $WW^*$ are proposed, which can be used to compare with the standard Higgs boson Lagrangian for these interactions.

In the event that the new boson is not a Higgs particle but a bound state $\zeta$ resonance, then we must consider an alternative mechanism that breaks electroweak symmetry. Many alternative models have been published. Three possible models have been proposed~\cite{Moffat3,Moffat4,Moffat5}. In the local renormalizable field theory model described in ref.~\cite{Moffat5}, there are no scalar boson modes and the masses of the $W$ and $Z$ bosons and fermions are produced by the quantum vacuum self-energies of the particles associated with a dynamical vacuum symmetry breaking mechanism.

\section*{Acknowledgements}

I thank Alvaro de Rujula and Viktor Toth for helpful discussions. This research was generously supported by the John Templeton Foundation and by the Perimeter Institute for Theoretical Physics. Research at the Perimeter Institute is supported by the Government of Canada through Industry Canada and by the Province of Ontario through the Ministry of Economic Development and Innovation.

\end{document}